\documentstyle[preprint,prd,aps,eqsecnum]{revtex}

\begin{document}
\hfill hep-th/9608065
\vskip 1cm 
\begin{center}
{\large \bf A Locally Supersymmetric Action for the Scalar Particle}
\vskip 2cm

{Fiorenzo Bastianelli\footnote{email: bastianelli@imoax1.unimo.it} and
Luca Consoli\footnote{email: consoli@imoax1.unimo.it}}
\vskip .8cm
{\sl Dipartimento di Fisica, Universit\'a di Modena, via Campi 213/A, 41100
Modena, Italy}
\vskip 4cm

{ Abstract}
\end{center}
We construct a locally supersymmetric action for the scalar particle, and study
its relation with the usual reparametrization invariant action. The mechanisms
at work are similar to those employed in the embedding of the bosonic string
into the fermionic one, originally due to Berkovits and Vafa in their search
for a universal string. The simpler algebraic structure present in the particle
case provides us with a guide on how to prove in a simple way, without the
need of fixing the superconformal gauge, that the supersymmetric formulation of
the bosonic string is equivalent to the usual one, where reparametrization
invariance is the only world-sheet gauge symmetry.

\vfill
\newpage

\section{Introduction}
\label{sec:intro}

String theory is still lacking a suitable non-perturbative formulation. The
best description available remains that of a perturbative expansion of strings
propagating on a consistent background. String coordinates are quantized, and
the different terms of the loop expansion are related to the different 
topologies of the world-sheet. 
In the absence of a more fundamental principle, different
string theories have been classified according either to their world-sheet
gauge symmetries or to their space-time properties. However, recent
developments have made it clear that these two ways of classifying strings have
not an intrinsic meaning; rather, they are seen to reflect properties of the
chosen vacuum. In fact, on one hand string theories classified according to
their space-time properties are related to each other by the so-called S, T and
U dualities. This suggests that they may be interpreted as different 
perturbative expansions of a single unifying theory (the so-called M
theory) about different vacua \cite{Schw}. Similarly, in the other
classification based on the world-sheet gauge symmetries (or, equivalently, on
the constraint algebra of the conformal gauge formulation), it has been shown
how the bosonic string described by a reparametrization invariant world-sheet 
action may be viewed as a special case of the fermionic string, 
which is described by a
locally supersymmetric world-sheet action \cite{Ber}. This hints on the
possible existence of a universal string theory containing all the other 
ones as 
particular broken symmetric phases. In fact, in ref. \cite{Bas1} a hierarchy 
of supersymmetric strings was discovered, in which each string of the
hierarchy contains also all those strings with a lower number of local
supersymmetries on the world-sheet.

In this article we study the mechanism of embedding, originally discovered by
Berkovits and Vafa, by applying it to a simpler case, namely the case of
particles. A scalar particle is described by an action with reparametrization
invariance, while the standard action for a spin-$\frac{1}{2}$ particle has a
local supersymmetry on the world-line \cite{Brin}. We show that it is
possible to construct a locally supersymmetric action describing the scalar
particle. We solve the corresponding 
master equation of Batalin and Vilkovisky
necessary for lagrangian quantization, and describe how this 
supersymmetric action is 
related to the usual reparametrization invariant one. 
We cast our discussion in the language of canonical 
transformations to exemplify that formalism. The
simpler algebraic setting present in the particle case suggests how to proceed
in the string case, and prove the equivalence of the locally supersymmetric
formulation with the usual reparametrization invariant one, without the need of
fixing the superconformal gauge. A simple field redefinition relates the
local supersymmetric action given in \cite{Bas2} to the standard formulation of
refs. \cite{Bri2}. This procedure bypasses subtleties of the superconformal
gauge related to the treatment of the moduli and supermoduli 
of higher genus Riemann surfaces \cite{Pol}. Eventually, we will present our 
conclusions.

\section{A rigid supersymmetric model}
\label{sec:sumo}

We repeat in a simpler context the construction presented in \cite{Ber}. In
that reference Berkovits and Vafa found a suitable superconformal system
that can be used as a background for the fermionic string in the
superconformal gauge. The fermionic string propagating on such a background
behaves exactly as the bosonic string. In a similar way, we will find a
supersymmetric quantum mechanics that will reproduce the behaviour of a scalar
particle once coupled to world-line supergravity. The supersymmetric action
defining the model is given by
\begin{equation}
  S=\int{d\tau\left(\frac{1}{2}\dot{x}^{\mu}\dot{x}_{\mu} +
  b_1\dot{c}_1\right),}
\label{act1}
\end{equation}
where $x^{\mu}$ are the coordinates of the particle, $\dot{x}^{\mu} =
\frac{d}{d\tau}x^{\mu}$, and $(b_1, c_1)$ are a pair of 
anticommuting variables that will
allow to realize supersymmetry. The rigid supersymmetry transformations are
given by
\begin{eqnarray}
  \delta x^{\mu} & = & \varepsilon c_1 \dot{x}^{\mu}, \nonumber \\
  \delta c_1 & = & \varepsilon + \varepsilon c_1\dot{c}_1 , \label{susybis} \\
  \delta b_1 & = & -\varepsilon\left(\frac{1}{2}\dot{x}^{\mu}\dot{x}_{\mu} +
  \dot{b}_1 c_1 + 2b_1\dot{c}_1\right). \nonumber
\end{eqnarray}
By computing the commutator algebra of two supersymmetry transformations one
obtains a translation in the time parameter.
This is the trademark of supersymmetry. 
The corresponding supersymmetry charge is
\begin{equation}
     Q = b_1 - \frac{1}{2}c_1\dot{x}^{\mu}\dot{x}_{\mu}.
\label{sc}
\end{equation}

We note that the first term in the transformation of $c_1$ ($\delta c_1
= \varepsilon +\ldots$) is the one that makes these 
rules different from the BRST rules arising after gauge-fixing the
standard bosonic particle action
\begin{equation}
   S=\int{d\tau\ \frac{1}{2}e^{-1}\dot{x}^\mu \dot x_\mu}
\label{actgf}
\end{equation}
with the gauge condition $e=1$, and identifying the fields $c_1$ and 
$b_1$ with the corresponding ghost and antighost, respectively. 
This fact is also seen in the term
proportional to $b_1$ appearing in the supersymmetry charge that makes it
different from the BRST charge
\begin{equation}
 Q_{BRST} = -\frac{1}{2}c_1\dot{x}^\mu \dot x_\mu.
\label{brst}
\end{equation}

Another observation is that the previous supersymmetry transformations 
can be simplified by adding suitable trivial symmetries proportional to the 
equations of motion, so to obtain
\begin{eqnarray}
 \delta x^{\mu} & = & \varepsilon c_1 \dot{x}^{\mu}, \nonumber \\
 \delta c_1 & = & \varepsilon, \label{susy1} \\
 \delta b_1 & = & -\frac{1}{2}\varepsilon\dot{x}^{\mu}\dot{x}_{\mu}. \nonumber
\end{eqnarray}
However, with these simplified rules
the supersymmetry algebra closes only on-shell on translations,
while the previous rules close nicely off-shell.

As a final remark, we note that the supersymmetric action in eq. (\ref{act1})
can be generalized to a
more general bosonic background described by a metric $g_{\mu\nu}(x)$, a vector
potential $A_{\mu}(x)$ and a scalar potential $V(x)$,
\begin{equation}
  S=\int{d\tau\left(\frac{1}{2}g_{\mu\nu}(x)\dot{x}^{\mu}\dot{x}^{\nu} +
  A_{\mu}(x)\dot{x}^{\mu} + V(x) + b_1\dot{c}_1\right)}.
\label{act2}
\end{equation}
The supersymmetry transformations are given now by
\begin{eqnarray}
  \delta x^{\mu} & = & \varepsilon c_1 \dot{x}^{\mu}, \nonumber \\
  \delta c_1 & = & \varepsilon + \varepsilon c_1\dot{c}_1, \label{susy2} \\
  \delta b_1 & = & -
\varepsilon\left(\frac{1}{2}g_{\mu\nu}(x)\dot{x}^{\mu}\dot{x}^{\nu} 
- V(x) + \dot{b}_1 c_1 + 2b_1\dot{c}_1\ \right). \nonumber
\end{eqnarray}
Quantizing this system will give a supersymmetric quantum mechanics of a
different type from the standard one \cite{Wit}.
However, it is a much less powerful supersymmetric
quantum mechanics, since it is realized on a non-positive definite
Hilbert space, being $(b_1, c_1)$ a system of ghost-like nature.
In fact, many standard results will not apply
(e.g. positive definiteness of the energy will not hold).
This will not be a problem for our purposes  
since we plan to couple the model to 
gauge fields, those of world-line supergravity, allowing
us to recover unitarity.

\section{A locally supersymmetric action for the scalar particle}
\label{sec:susy}

The system in (\ref{act1}) can be coupled to world-line supergravity. This
will give a locally supersymmetric action, i.e. an action with the same local
symmetries of the one describing the spin-$\frac{1}{2}$ particle. Using the
Noether method to gauge the global symmetry in eq. (\ref{susybis}), 
and dropping the space-time indices on the coordinates,
we obtain
\begin{equation}
 S_1 =\int{d\tau\left(\frac{1}{2}e^{-1}\dot{x}^2 + b_1\dot{c}_1 + \chi b_1
   -\frac{1}{2}e^{-2}\chi c_1\dot{x}^2\right)},
\label{act3}
\end{equation}
where $e$ and $\chi$ are the einbein and the gravitino fields of the world-line,
respectively. The local supersymmetry transformations are given by
\begin{eqnarray}
 \delta_Q x & = & \varepsilon e^{-1} c_1 \dot{x}, \nonumber \\
 \delta_Q c_1 & = & \varepsilon + \varepsilon e^{-1}
c_1 (\dot{c}_1 - \chi) ,\nonumber \\
 \delta_Q b_1 & = & -\frac{1}{2}\varepsilon e^{-2}(1 - e^{-1}\chi
c_1)\dot{x}^2  - \varepsilon  e^{-1} \dot{b}_1 c_1   
- 2 \varepsilon e^{-1} b_1 (\dot{c}_1-\chi) , \label{susy3} \\
 \delta_Q e & = & 2\varepsilon\chi ,\nonumber \\
 \delta_Q \chi & = & \dot{\varepsilon} ,\nonumber
\end{eqnarray}
while the reparametrizations are given by
\begin{eqnarray}
  \delta_R x & = & \xi\dot{x} ,\nonumber \\
  \delta_R c_1 & = & \xi\dot{c_1} ,\nonumber \\
  \delta_R b_1 & = & \xi\dot{b}_1 ,\label{rep1} \\
  \delta_R e & = & \partial_{\tau}{(\xi e) } ,\nonumber \\
  \delta_R \chi & = & \partial_{\tau}{(\xi\chi)}. \nonumber
\end{eqnarray}
The transformation rules on the supergravity multiplet $(e,\chi)$ are the
standard ones \cite{Brin}, and the algebra of local symmetries closes 
off-shell on all the fields
\begin{equation}
 \begin{array}{rcl}
\, [\delta_Q(\varepsilon_1),\delta_Q(\varepsilon_2)]& = &
\delta_R(\xi=2\varepsilon_2\varepsilon_1 e^{-1}) +
\delta_Q(\varepsilon = -2\varepsilon_2\varepsilon_1 e^{-1}\chi) ,\\
   \,[\delta_R(\xi_1),\delta_R(\xi_2)] & = & \delta_R(\xi = \xi_2\dot{\xi}_1 -
\xi_1\dot{\xi}_2) ,\\
   \,[\delta_R(\xi_1),\delta_Q(\varepsilon_2)] & = & \delta_Q(\varepsilon =
-\xi_1\dot{\varepsilon_2}).
 \end{array}
\label{alg3}
\end{equation}

Because the complete algebra closes off-shell, it is straightforward to solve
the Batalin-Vilkovisky master equation, useful for 
gauge-fixing in the lagrangian quantization \cite{Gom}. 
The proper solution is given by
\begin{eqnarray}
 S_{1, BV} &=& S_1 + 
\int d\tau \biggl[ x^* (\eta - \gamma e^{-1}c_1) \dot{x}
+ c_1^* \biggl( \dot c_1 \eta + \gamma +\gamma e^{-1}c_1 (\dot c_1 -\chi)
\biggr) 
\nonumber\\
&+& b_1^* \left( \dot b_1 \eta -\frac{1}{2} \gamma e^{-2} (1- e^{-1} \chi c_1)
\dot{x}^2 - \gamma e^{-1}\dot b_1 c_1 - 2 \gamma e^{-1}
b_1 (\dot c_1 -\chi) \right)  \label{actBVuno}\\
&+& e^*(\partial_{\tau}(\eta e) - 2 \gamma \chi) 
+ \dot \chi^* ( \eta\chi - \gamma )
+ \eta^* (\dot\eta \eta +\gamma^2 e^{-1})
+ \gamma^* (\eta\dot\gamma -\gamma^2 e^{-1}\chi) \biggr],
  \nonumber
\end{eqnarray}
where $\eta$ and $\gamma$ are the ghosts for $\xi$ and
$\varepsilon$, respectively, and the starred fields are the so-called
antifields (i.e. the sources for the BRST variations).

To see why this action describes a scalar particle, we note
that variables can be redefined and that the basis of gauge symmetries is
not unique, since it can be modified considerably by adding trivial gauge
symmetries proportional to the equations of motions.
In fact, after a little inspection, we see that by defining the new variables
\begin{equation}
\begin{array}{rcl}
\tilde{e} & = & e + \chi c_1 ,\\
\tilde{\chi} & =& \chi -\dot{c}_1 ,
\end{array}
\label{ridef1}
\end{equation}
the action (\ref{act3}) can be rewritten as follows
\begin{equation}
   S_2 = \int{d\tau\left(\frac{1}{2}\tilde{e}^{-1}\dot{x}^2 +
        \tilde{\chi} b_1\right)}.
\label{act4}
\end{equation}
Moreover, the two gauge symmetries can also be represented by
\begin{eqnarray}
 \delta x & = & \tilde{\xi}\dot{x} ,\nonumber \\
 \delta c_1 & = & \tilde{\varepsilon} ,\nonumber \\
 \delta b_1 & = & 0 ,\label{rep2} \\
 \delta \tilde{e} & = & \partial_{\tau}{(\tilde{\xi}\tilde{e})} ,\nonumber \\
 \delta \tilde{\chi} & = & 0 ,\nonumber
\end{eqnarray}
where $\tilde{\xi}$ and $\tilde{\varepsilon}$ are the two independent gauge
parameters. We see immediately that
the variable $c_1$ can be dropped since it is a pure gauge degree
of freedom, while the variables $b_1$ and $\tilde{\chi}$ are non-dynamical
fields which can be eliminated by their equations of motion. Thus, we are left
with the standard reparametrization invariant action describing a massless
scalar particle, and quantization can proceed in the well-known way.
This proves our claim that a scalar particle can be described
by a locally supersymmetric action. Note that the gauge symmetries
in eqs.
(\ref{susy3}) and (\ref{rep1}), when expressed in terms of the new
variables, can be written as follows
\begin{eqnarray}
  \delta x & = & \tilde \xi \dot{x} ,\nonumber \\
  \delta c_1 & = & \tilde \varepsilon + \tilde \xi\dot{c}_1 ,\nonumber \\
  \delta b_1 & = & \tilde{\xi}\partial_{\tau}{\frac{\delta_L S_2}{\delta 
       \tilde{\chi}}} +
       \tilde{\varepsilon}\displaystyle\frac{\delta S_2}{\delta\tilde{e}} 
       -2\tilde{\varepsilon}\tilde{e}^{-1}b_1{\frac{\delta_L S_2}{\delta b_1}}
       \label{gauge} ,\\
  \delta \tilde{e} & = & \partial_{\tau}(\tilde{\xi}\tilde{e}) 
               - \tilde{\varepsilon}
           \displaystyle\frac{\delta_L S_2}{\delta b_1} ,\nonumber \\
  \delta \tilde{\chi} & = & -\partial_{\tau}{\left(
        \tilde{\xi}\frac{\delta_L S_2}{\delta b_1}\right)} ,\nonumber 
\end{eqnarray}
where we have defined for convenience
\begin{equation}
 \begin{array}{rcl}
\tilde{\xi} & = & \xi + \varepsilon e^{-1} c_1 ,\\
   \tilde{\varepsilon} & = & \varepsilon -\varepsilon e^{-1} c_1 \chi ,
 \end{array}
\label{ridef3}
\end{equation}
and where ${\delta_L S \over \delta \phi}$ denotes the
left functional derivative of the action.
Dropping the terms proportional to the equations of motion from the right 
hand side, and shifting $\tilde{\varepsilon} \to
\tilde{\varepsilon} - \tilde{\xi} \dot{c}_1$
one obtains the simplified basis of gauge symmetries.
Thus, we see how the simple gauge symmetries in eq. (\ref{rep2}),
containing the reparametrizations and a shift symmetry on a fermionic 
variable, can mimic the full 
transformations of local supersymmetry.

The proper solution of the master equation for the action in (\ref{act4}) with
gauge symmetries in (\ref{rep2}) can be obtained quite easily.
Dropping the tilde on the variables we have
\begin{equation}
 S_{2, BV} =  \int{d\tau\left( \frac{1}{2}e^{-1}\dot{x}^2 + \chi b_1
 + x^*\eta\dot{x} + 
e^*\partial_{\tau}(\eta e) + c_1^*\gamma + \eta^*\dot{\eta}\eta\right)} .
\label{actBV}
\end{equation}

The actions $S_{1, BV}$ and $ S_{2, BV}$ describe the same model and are
both proper solutions of the master equation. A theorem guarantees that 
these actions must be related by canonical transformations. 
In the
Batalin-Vilkovisky formalism, canonical transformations are typically used to
gauge-fix and to redefine variables (in fact, the process of gauge-fixing can
be thought of as a particular field redefinition). They preserve the properties
of the antibracket and map proper solutions into proper solutions. Canonical
transformations are specified by a fermionic generating function $\Psi$ and are
defined by $\phi \rightarrow \phi^{\prime} = e^{{\cal{L}}_{\Psi}}\phi$, where
$\phi$ is a field or an antifield, and ${\cal L}_{\Psi}\phi\equiv (\Psi,\phi)$
with $(\ ,\ )$ denoting the antibracket.
The canonical transformation 
between $S_{1, BV}$ and $ S_{2, BV}$ can be presented in a factorized form
\begin{equation}
S_{2, BV} = e^{{\cal L}_{\Psi_7}} e^{{\cal L}_{\Psi_6}} 
e^{{\cal L}_{\Psi_5}} e^{{\cal L}_{\Psi_4}}
e^{{\cal L}_{\Psi_3}} e^{{\cal L}_{\Psi_2}}
e^{{\cal L}_{\Psi_1}} S_{1, BV}
\end{equation}
where the various gauge fermions are given by
\begin{eqnarray}
 \Psi_1 & = & \int{d\tau\ e^*\chi c_1} ,\nonumber \\
 \Psi_2 & = & - \int{d\tau\ \chi^*\dot{c_1}} ,\nonumber \\ 
 \Psi_3 & = & -\int{d\tau\ \eta^*\gamma e^{-1} c_1 } ,\nonumber \\
 \Psi_4 & = & -\int{d\tau\ \gamma^*\gamma e^{-1}c_1(\dot{c_1}+\chi)} ,
\label{catr} \\
 \Psi_5 & = & \int{d\tau\ \biggl[e^* \chi c_1 + b_1^* c_1 \biggl(\frac{1}{2}
 e^{-2} \dot{x}^2 - 2 e^{-1} b_1 \chi\biggr)\biggr] } ,\nonumber \\
 \Psi_6 & = & -\int{d\tau\  \dot{\chi}^*b_1^*\eta} ,\nonumber \\
 \Psi_7 & = & -\int{d\tau\ \gamma^*\eta\dot{c_1}} .\nonumber
\end{eqnarray}
We note that the transformations specified by $\Psi_1$ and $\Psi_2$ 
accomplish the field redefinition given in eq. (\ref{ridef1}),
$\Psi_3$ and $\Psi_4$ redefine the ghosts (i.e. the gauge parameters)
as in eq. (\ref{ridef3}), $\Psi_5$ and $\Psi_6$
modify the gauge symmetry basis by terms proportional to the equations
of motion as in eq. (\ref{gauge}), and $\Psi_7$
performs a final redefinition of the ghost $\gamma$.

In ref. \cite{Bas2}, the same method of canonical transformations was used 
to prove the equivalence between the fermionic string propagating on the 
Berkovits-Vafa background and the bosonic string.
However, that proof was given in the conformal gauge, as 
was the original proof in \cite{Ber} and the conformal 
field theoretical proof in \cite{Ish}.
Here, in the simpler case of particles, we have been able
to treat the geometrical fields on the world-line, the einbein and 
the gravitino,
in their full generality, without the need of specifying a gauge condition.
The lesson learned in this simple model will help us in re-examining the 
string case from a better perspective.

Before closing this section, we note that it is straightforward to 
generalize the  supersymmetric action for the particle by including
a mass term, and, in general, the full background
described in eq. (\ref{act2}) (a constant term in the potential $V$ can
reproduce the effect of a mass term for the particle).
For simplicity we have discussed here only 
the basic case of a massless particle propagating on a flat space-time.

\section{Equivalence of the fermionic string on the Berkovits-Vafa background
and the bosonic string}
\label{sec:equiv}

We have seen in the previous section how a locally supersymmetric formulation
of the scalar particle can be related to the standard reparametrization
invariant one. The relation is given by a field redefinition consisting 
in a shift of the gravitino field, and in a certain transformation on the
einbein. The latter can also 
be interpreted as a particular supersymmetry transformation.
Guided by this particular field redefinition, we will relate the
locally supersymmetric formulation of the bosonic string to the usual
reparametrization invariant one. The locally supersymmetric action, which in 
the superconformal gauge reproduces the model of Berkovits and Vafa, was
constructed in ref. \cite{Bas2}, and reads:
\begin{eqnarray}
 S = \frac{1}{\pi}\int d^2 x e \biggl (&\frac{1}{2}&
\nabla_{_{\!\!+\!\!\!+}} X \nabla_{_{\!\!=}} X +
b_1\nabla_{_{\!\!=}}c_1 + \bar{b}_1\nabla_{_{\!\!+\!\!\!+}}\bar{c}_1 + 
\chi_{_{-=}}G_{_{++\!\!\!+}} +
     \chi_{_{++\!\!\!+}}G_{_{-=}}  \label{action1} \\
 &+& \chi_{_{++\!\!\!+}} \chi_{_{-=}}c_1\bar{c}_1
\nabla_{_{\!\!+\!\!\!+}} X \nabla_{_{\!\!=}} X \biggr ),  
     \nonumber
\end{eqnarray}
where
\begin{equation}
\begin{array}{rcl}
 G_{_{++\!\!\!+}} & = & b_1(1+c_1\nabla_{_{\!\!+\!\!\!+}}c_1) -
\frac{1}{2}c_1(\nabla_{_{\!\!+\!\!\!+}}X)^2 , \\
 G_{_{-=}} & = & \bar{b}_1(1+\bar{c}_1\nabla_{_{\!\!=}}\bar{c}_1)
 -\frac{1}{2}\bar{c}_1(\nabla_{_{\!\!=}}X)^2 ,\nonumber 
\end{array}
\label{defi}
\end{equation}
and where $(b_1, c_1, \bar{b}_1, \bar{c}_1)$ are Lorentz tensors with spin
$(\frac{3}{2}, -\frac{1}{2}, -\frac{3}{2}, \frac{1}{2})$, $\nabla_a =
{e_a}^{\mu}\partial_{\mu} + \omega_a J$ is the Lorentz covariant derivative
with the flat index $a$ taking values $(+\!\!\!+,=)$, 
$J$ is the Lorentz generator
which measures the Lorentz spin, and $\omega_a$ is the spin connection (see
ref. \cite{Bas2} for more extended definitions).
The following redefinition of variables 
\begin{equation}
 \begin{array}{rcl}
  {\tilde{e}}_{_{+\!\!\!+}}{}^{\mu} & = & {e_{_{+\!\!\!+}}}^{\mu} - 
  \chi_{_{++\!\!\!+}}\bar{c}_1{e_{_=}}^{\mu} ,\\
  {\tilde{e}_{_=}}{}^{\mu} & = & {e_{_=}}^{\mu} - 
\chi_{_{-=}}c_1{e_{_{+\!\!\!+}}}^{\mu} ,\\
  \tilde{\chi}_{_{++\!\!\!+}} & = & e(\chi_{_{++\!\!\!+}} +
  \chi_{_{++\!\!\!+}}\bar{c}_1\nabla_{_{\!\!=}}\bar{c}_1 - 
\nabla_{_{\!\!+\!\!\!+}}\bar{c}_1) ,\\
  \tilde{\chi}_{_{-=}} & = & e(\chi_{_{-=}} + \chi_{_{-=}}c_1
\nabla_{_{\!\!+\!\!\!+}}
c_1 - \nabla_{_{\!\!=}}c_1) ,
 \end{array}
\label{ridef2}
\end{equation}
brings the action (\ref{action1}) in the form
\begin{equation}
  S = \frac{1}{\pi}\int{
d^2x\left(\frac{\tilde{e}}{2}\tilde{\nabla}_{_{\!\!+\!\!\!+}}
X\tilde{\nabla}_{_{\!\!=}}X 
+ \tilde \chi_{_{-=}} b_1 +\tilde \chi_{_{++\!\!\!+}}\bar{b}_1
\right)} .
\label{action2}
\end{equation}
Once more, the fields $(c_1, \bar{c}_1)$ can be dropped since 
the action does not depend on them, and so they are pure
gauge degrees of freedom, while $(b_1, \tilde{\chi}_{_{-=}})$ and $(\bar{b}_1,
\tilde{\chi}_{_{++\!\!\!+}})$ are non-dynamical auxiliary fields which can be
eliminated by their equations of motion. This leaves us with the standard
action for the bosonic string, written here
using the world-sheet vielbein rather than the world-sheet metric.
At this point, 
one could  also discuss how the supersymmetry 
transformations for the action (\ref{action1}), given in \cite{Bas2},
are mimicked  by the natural gauge symmetries of 
eq.  (\ref{action2}), reparametrizations and local shifts in 
$(c_1 ,\bar c_1)$. However, we leave this as an exercise,
since it is quite similar to the particle case discussed above.

The advantage of our proof of the equivalence between the fermionic string 
on the Berkovits-Vafa background and the bosonic string
is that we didn't have to fix the conformal gauge.
The latter is typically too strong a condition on the geometrical fields
of the world-sheet, and modular integrations corresponding to the 
antighosts zero modes must be performed.
We recognize that, after fixing the conformal gauge
for the supersymmetric model discussed above, one would get  
integrations over the moduli corresponding to the gravitino, but these 
would be just  gauge artifacts. Our proof dispenses us from
analyzing how the particular mechanism for these modular integrations 
would work out.

\section{Conclusions}
\label{sec:conclu}

We have constructed and analyzed a locally supersymmetric 
formulation for the scalar particle. This model employes in a simple
context the mechanism discovered by Berkovits and Vafa for embedding the
bosonic string into the fermionic one.
Guided by the particle model we have been able to give a simpler proof of
the string embedding.
String theory is supposed to determine its own background,
so such an embedding is taken as an indication of the existence of
a \lq\lq universal string theory".
However, this is a non-perturbative statement, and very little
information can be extracted by the existence of the above embedding.
In fact, as we have seen, properties arising from such embeddings are 
essentially gauge artifacts. 
In contrast, the other hint about the existence of a unifying string theory,
which comes from S, T and U dualities, has been much more fruitful, 
since it has a lot of predictive power \cite{Schw}.
One should remember that any gauge algebra
can always be brought into an abelian form \cite{Bata}. 
Sometimes, this clashes with manifest locality and/or manifest 
covariance with respect to certain symmetries  
(like Lorentz covariance in standard Yang-Mills theory),
and thus the non-abelian basis may be a preferred one.
As we have seen,
in the first quantized description of particles and strings,
there does not seem to exist a preferred basis for the gauge symmetries
that can be used for an intrinsic definition of the various models.

\end{document}